# The two-positron gluic bond as a manifestation of 'super' van der Waals interactions

Mohammad Goli[1], Dario Bressanini[2], Shant Shahbazian[3]

[1]Institute of Physics, Faculty of Physics, Astronomy, and Informatics, Nicolaus Copernicus University in Toruń, ul. Grudziądzka 5, 87-100 Toruń, Poland,
E-mail: mgoli@umk.pl

[2]Dipartimento di Scienza e Alta Tecnologia, Università dell'Insubria, Como, Italy
E-mail: dario.bressanini@uninsubria.it

[3]Alikhanian National Laboratory (Yerevan Physics Institute), Alikhanian Brothers Street 2, Yerevan 0036, Armenia
E-mail: shant.shahbazyan@aanl.am

## Abstract

Recently, it has been demonstrated theoretically that the interaction of two $PsH$ atoms — each being a stable bound state of a hydrogen atom and a positronium atom — is attractive, leading to the formation of a molecular complex denoted as $(PsH)_2$. However, the physical nature of this interaction has remained elusive. In the present study, we show that the stabilizing mechanism is entirely encoded in the quantum correlations between the two positrons and, to a lesser extent, in the electron-positron correlations. Notably, the interaction cannot be recovered at the mean-field (Hartree-Fock) level, nor by computational models that include only electron-electron correlation effects. Accordingly, the bond formed between $PsH$ units — termed here a *two-positron gluic bond* to emphasize its fundamentally distinct character from the two-positron covalent bonds present in pure antimatter molecules — emerges only when matter and antimatter particles form a common bound state. When classified within the framework of known bonding mechanisms, this gluic bond falls into the category of stabilizing dispersion interactions, giving rise to a van der Waals complex. However, its remarkably large bond dissociation energy, compared with those of strongly bonded van der Waals complexes of similar size, reveals an anomalously strong interaction. For this reason, we propose that $(PsH)_2$ is most appropriately described as a *"super" van der Waals complex* stabilized by a *"super" van der Waals bond*.

## Introduction

The concept of the two-electron homopolar bond was first introduced in Lewis's ad hoc atomic theory and later gained a solid foundation through the quantum mechanical analysis of the hydrogen molecule by Heitler and London (HL).[1–3] Since then, the two-electron covalent bond has remained central to both quantitative and qualitative understanding of molecular bonding,[4–6] and has also shaped our understanding of condensed phases.[7–9] The analog of this bond in an antimatter world, wherein antiparticles replace particles, is the two-positron covalent bond. With recent advances in the production, trapping, and cooling of the antihydrogen atoms,[10–12] experimental realization of the antihydrogen molecule—and the accompanying two-positron bond—appears inevitable.[13] If the symmetry holds exactly or is even only weakly violated between hydrogen and antihydrogen atoms,[14] a two-positron covalent bond is indistinguishable from its electronic analogue. In that case, more generally, chemistry and antichemistry would exist as exact mirror images. However, extending beyond pure matter or antimatter systems to hybrid matter-antimatter mixtures reveals the emergence of entirely novel positron bonds.[15–24]

Molecules composed of both matter and antimatter are inherently short-lived, being intrinsically unstable to matter–antimatter annihilation. Nevertheless, intramolecular bonding can provide stability against specific decomposition channels into atomic fragments. Reyes and coworkers first demonstrated computationally that a single positron may bind two hydrides (denoted as $\left[H^-,e^+,H^-\right]$),[15] or halides,[25] closing all the considered decomposition channels. These anions repel each other in the absence of the positron and do not form stable species; thus, the formation of the one-positron bond is the sole origin of binding. The analysis of the positron density in $\left[H^-,e^+,H^-\right]$ revealed that it is mainly concentrated in the middle of the two protons, acting as a positively charged glue to bind the two negatively charged hydrides.[15,16] Also, a detailed energy decomposition analysis revealed a relatively simple bonding mechanism in this species through electrostatic interactions, resembling a conventional ionic bond in which two oppositely charged ions attract electrostatically.[16] In effect, the positron behaves like a 'virtual cation' positioned between the two anions. This type of bond seems to be peculiar to positron-carrying species, and it has been called a one-positron "gluonic" bond to make it distinct from the usual positronic covalent bonds in the purely antimatter molecules (in response to feedback and to avoid possible confusion with the established particle-physics meaning of the term gluon, we rename this bond to "gluic" bond).[22]

Adding another positron to $\left[H^{-},e^{+},H^{-}\right]$ yields a new species, which has been computationally shown to be stable against various dissociation channels.[19] Subsequent analysis of the two-positron gluic bond in this species revealed that it is more complex than the one-positron bond.[22] In this species, the positron density is also concentrated in the middle of the two hydrides, implying an electrostatic attractive interaction between positrons and the hydrides. However, the energy decomposition analysis reveals that this attraction is insufficient for stabilization since the additional positron contributes to stabilizing interactions while simultaneously introducing destabilizing effects absent in $\left[H^{-},e^{+},H^{-}\right]$.[22] Thus, another stabilizing mechanism seems to be operative in this species, as considered in the present study. A detailed analysis of the radial distribution function of various pairs of particles revealed that this species is a mildly perturbed combination of two "exotic" $PsH$ atoms and denoted as $\left(PsH\right)_2$;[22] although it is an atom, $PsH$ is itself a combination of hydrogen, $H$, and positronium, $Ps$,[26] atoms. The stability of $PsH$ with respect to dissociation into its constituents was predicted long ago,[27] and since then, it has been produced and studied experimentally.[28] The two-positron gluic bond is formed between two charge-neutral $PsH$ atoms, and in the present study, the aim is to elucidate the nature of this bond, which, as is demonstrated, is rooted in the quantum correlations among the positrons, as well as between electrons and positrons.

**Results and discussion**

The distinction between one- and two-positron gluic bonds gives a primary clue to the unique nature of the latter. For the potential energy curve (PEC) of $\left[H^-,e^+,H^-\right]$, derived from the fully correlated diffusion quantum Monte Carlo method (DMC),[29–31] (Table S1 in supporting information (SI)), the equilibrium bond length, $R_e$, and bond dissociation energy (BDE), $BDE = E_{H^-}^{DMC}\left(R_e\right) + E_{PsH}^{DMC}\left(R_e\right) - E_{\left[H^-,e^+,H^-\right]}^{DMC}\left(R_e\right)$, are around 6.3 Bohr and 62 kJ.mol$^{-1}$ (the zero-point corrected BDE is almost 59 kJ.mol$^{-1}$), respectively. At the least correlated computational level, i.e. the multi-component Hartree-Fock method (MC-HF),[32] employing the aug-cc-pVQZ hydrogenic basis set for both positronic and electronic spatial orbitals,[33,34] (MC-HF/[QZ:QZ]), the corresponding values are around 7.2 Bohr and 60 kJ.mol$^{-1}$, (see Table S1 in the SI). At first glance, the difference of almost 0.9 Bohr between the computed $R_e$ at these two levels may appear substantial. However, this impression is misleading due to the flatness of the PECs between 6.3 and 7.2 Bohr. At 6.3 Bohr, the correlation energy, $E_{\left[H^-,e^+,H^-\right]}^{corr} = E_{\left[H^-,e^+,H^-\right]}^{MC-HF} - E_{\left[H^-,e^+,H^-\right]}^{DMC}$, contributes only 7 kJ.mol$^{-1}$ to the BDE, and MC-HF/[QZ:QZ] energy recovers almost 88% of the BDE. These findings indicate that the primary driving force behind the formation of the one-positron gluic bond is not correlation-related effects. Structural analysis using the multi-component quantum theory of atoms in molecules (MC-QTAIM) partitioning methodology,[35,36] revealed a homonuclear system composed of two equivalent atoms, each centered on a proton and containing, on average, two electrons and half of the positron populations.[16] The one-positron bond is formed between these two net negatively charged atoms. A more detailed investigation, employing the two-component interacting quantum atoms (TC-IQA) energy decomposition analysis,[16] pinpointed the bond's origin to the electrostatic stabilization arising from the interaction between the positron's charge density associated with one atom and the electron's charge density associated with the other. This is the "classical" part of the gluic bonds, which is solely responsible for bonding in $\left[H^{-},e^{+},H^{-}\right]$.

In contrast, the DMC-derived PEC of $\left(PsH\right)_2$, (Figure 1 and Table S2 in the SI), yields $R_e$ of almost 6.0 Bohr and a BDE, $BDE=2\times E_{PsH}^{DMC}\left(R_e\right)-E_{(PsH)}^{DMC}\left(R_e\right)$, of 27 kJ.mol$^{-1}$ (the zero-point-corrected BDE is around 24 kJ.mol$^{-1}$). The 0.3 Bohr difference in $R_e$ compared to $\left[H^{-},e^{+},H^{-}\right]$, has no physical significance due to the extreme flatness of PECs near $R_e$. More strikingly, however, the BDE of $\left(PsH\right)_2$ is less than half that of $\left[H^{-},e^{+},H^{-}\right]$, indicating that the addition of the second positron substantially weakens the bond. At the MC-HF/[QZ:QZ] level, the computed $R_e$ and BDE are 8.3 Bohr and 1 kJ.mol$^{-1}$, respectively. The corresponding PEC at this level (Figure 1 and Table S2 in the SI) is again very flat in the range of 7.6 to 9.2 Bohr, and the very small BDE suggests the absence of a genuine local minimum. At 6.0 Bohr, the correlation energy contributes almost 35 kJ.mol$^{-1}$ to the BDE, whereas the MC-HF/[QZ:QZ] energy contributes negatively, -8 kJ.mol$^{-1}$, thereby reducing the overall BDE. These findings demonstrate that, in sharp contrast to the one-positron bond, the two-positron gluic bond is driven by correlation-related effects. Structural analysis using the MC-QTAIM revealed a homonuclear diatomic system composed of two slightly "deformed", charge-neutral $PsH$ atoms.[22] Application of the TC-IQA energy decomposition analysis showed that the same stabilizing mechanism identified in $\left[H^-,e^+,H^-\right]$ also operates in $\left(PsH\right)_2$ and indeed constitutes the only major stabilizing interaction between the two atoms at the MC-HF/[QZ:QZ] level.[22]

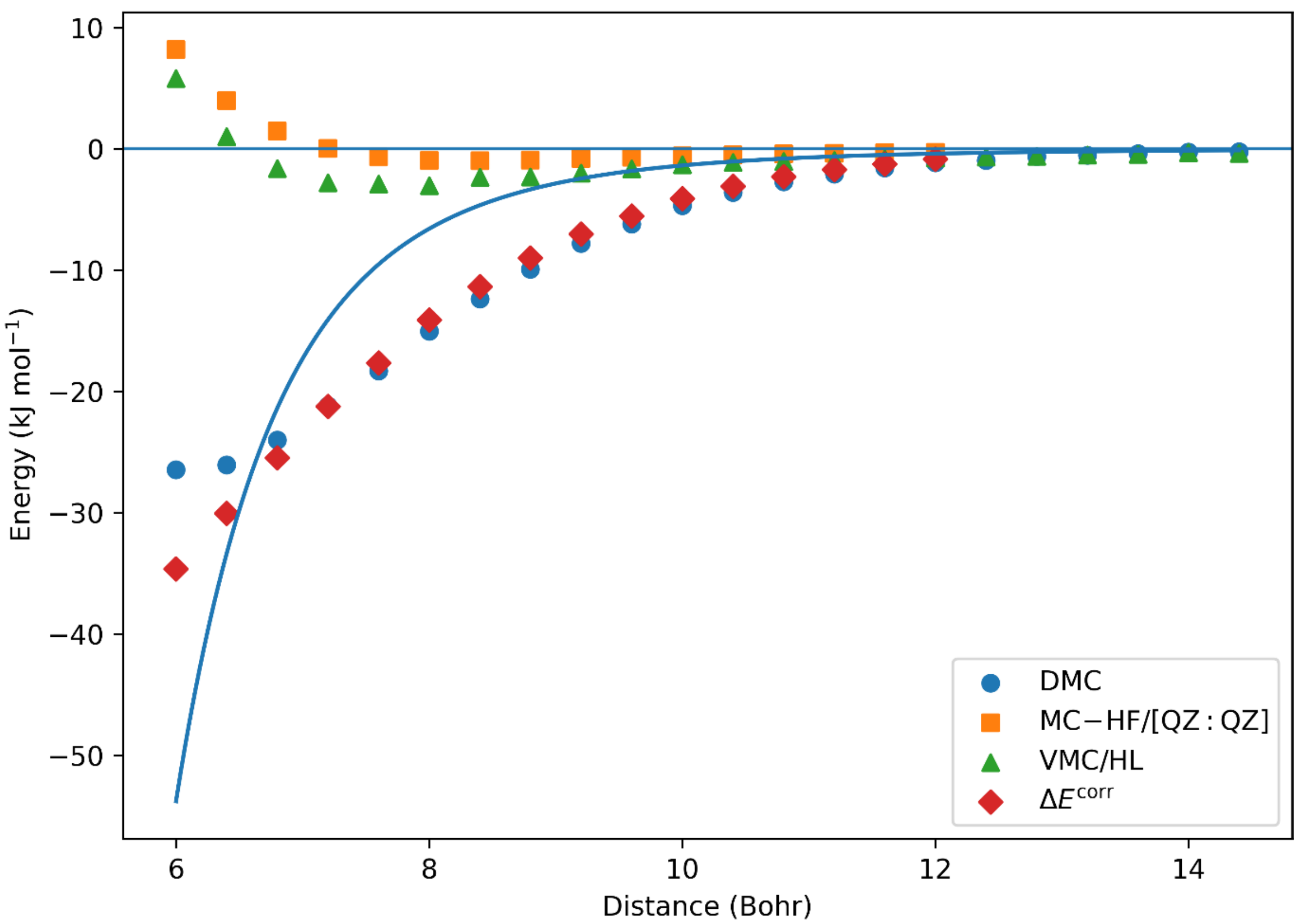

**Figure 1**- The potential energy curves (PECs) of $\left(PsH\right)_2$ computed at the DMC, MC-HF/[QZ:QZ], VMC/HL levels together with the calculated interatomic correlation energy, $\Delta E^{corr}\left(R\right)$, as functions of the inter-proton distance (see the text for definitions and abbreviations). For clarity and ease of comparison, the portion of the PECs on the left-hand side of the equilibrium distance, $R_e = 6$ Bohr, is not shown. The blue curve represents the estimated dispersion energy, $E_{dispersion}^{PsH-PsH}$ (see the text for the employed expression).

The key question, then, is why this interaction fails to provide sufficient binding to stabilize $\left(PsH\right)_2$ at the MC-HF level? To address this issue, the contributions of each fundamental operator in the Hamiltonian of the two systems were disentangled by analyzing their shares in DEs at two inter-proton distances, $DE(R)=2\times E_{PsH}^{DMC}(R)-E_{(PsH)_2}^{DMC}(R)$, with the results summarized in Table 1.

Table 1- The contribution of the kinetic, $\hat{T}_e$ and $\hat{T}_p$, and the Coulombic interaction potential energy operators, $\hat{V}_{e-nuc}$, $\hat{V}_{ee}$, $\hat{V}_{nuc-nuc}$, $\hat{V}_{p-nuc}$, $\hat{V}_{pp}$ and $\hat{V}_{ep}$, to the dissociation energies (DE) of $\left[H^-,e^+,H^-\right]$ and $(PsH)_2$, computed at 6 and 8 Bohr inter-proton distances (all given in kJ.mol$^{-1}$).*

| ***R* = 6.0** | $\Delta\langle\hat{T}_e\rangle$ | $\Delta\langle\hat{V}_{e-nuc}\rangle$ | $\Delta\langle\hat{V}_{ee}\rangle$ | $\Delta\langle\hat{V}_{nuc-nuc}\rangle$ | $\Delta\langle\hat{T}_p\rangle$ | $\Delta\langle\hat{V}_{p-nuc}\rangle$ | $\Delta\langle\hat{V}_{pp}\rangle$ | $\Delta\langle\hat{V}_{ep}\rangle$ | *DE* |
|---|---|---|---|---|---|---|---|---|---|
| $\left[H^-,e^+,H^-\right]$ | -141 | 1930 | -1785 | -438 | -51 | -603 | 0 | 1136 | 49 |
| $(PsH)_2$ | -75 | 1813 | -1740 | -438 | 4 | -886 | -427 | 1741 | -8 |
| ***R* = 8.0** | | | | | | | | | |
| $\left[H^-,e^+,H^-\right]$ | -21 | 1372 | -1357 | -328 | -16 | -379 | 0 | 785 | 56 |
| $(PsH)_2$ | -10 | 1323 | -1314 | -328 | 5 | -655 | -330 | 1311 | 1 |

*The subscripts *e*, *p*, and *nuc* stand for electron, positron, and nucleus (proton), respectively. Each entry has been calculated as follows: $\Delta\langle\hat{X}\rangle=\langle\hat{X}\rangle_A+\langle\hat{X}\rangle_B-\langle\hat{X}\rangle_{AB}$, where $\langle\hat{X}\rangle$ is the expectation of operator $\hat{X}$, computed employing MC-HF/[QZ:QZ] wavefunction. For DE, $\hat{X}=\hat{H}$, where $\hat{H}=\hat{T}_e+\hat{T}_p+\hat{V}_{e-nuc}+\hat{V}_{ee}+\hat{V}_{nuc-nuc}+\hat{V}_{p-nuc}+\hat{V}_{pp}+\hat{V}_{ep}$ is the Hamiltonian of the system; BDE is the DE computed at the corresponding $R_e$. The subscript *AB* represents the molecules, and the subscripts *A* and *B* stand for the atoms produced in the dissociation limit; in the case of $\left[H^-,e^+,H^-\right]$, $A=PsH, B=H^-$ and in the case of $(PsH)_2$, $A=B=PsH$.

It is straightforward to verify that, although the mean values of individual operators differ between the two species (except for $\Delta\langle\hat{V}_{nuc-nuc}\rangle$ since the systems are being compared at the equal inter-proton distances), the sum of the purely electronic contributions, $\Delta\langle\hat{T}_e\rangle+\Delta\langle\hat{V}_{e-nuc}\rangle+\Delta\langle\hat{V}_{ee}\rangle+\Delta\langle\hat{V}_{nuc-nuc}\rangle$, is remarkably similar. At 6 (8) Bohr, this sum amounts to almost -433 (-334) kJ.mol$^{-1}$ for $\left[H^-,e^+,H^-\right]$ and -440 (-330) kJ.mol$^{-1}$ for $(PsH)_2$. Among the four remaining terms, the only one with a major positive contribution to the DE, is $\Delta\langle\hat{V}_{ep}\rangle$. This stabilizing interaction represents the origin of the classical gluic bonding mechanism when dissected properly into atomic and inter-atomic contributions, through the TC-IQA, as discussed elsewhere.[16,22] Notably, the addition of a second positron to $\left[H^-,e^+,H^-\right]$ significantly enhances the magnitude of $\Delta\langle\hat{V}_{ep}\rangle$ at both considered inter-proton distances. At the same time, however, the destabilizing contribution from $\Delta\langle\hat{V}_{p-nuc}\rangle$ also increases, and an additional destabilizing term, $\Delta\langle\hat{V}_{pp}\rangle$, emerges. The delicate balance among these competing effects – resisting a simple qualitative analysis – ultimately determines the very small BDE of $(PsH)_2$ at the MC-HF level.

In the next step, to assess the role of correlations, a partially correlated state was considered, which was designed to further elucidate the nature of bonding in $(PsH)_2$. The PEC was computed for the spin-singlet HL wavefunction, originally developed for the hydrogen molecule,[1,2] $\Psi^{HL}_{(PsH)_2}(1,2)=\frac{1}{\sqrt{2(1+S^2_{AB})}}\left(\psi^A_{PsH}(1)\psi^B_{PsH}(2)+\psi^A_{PsH}(2)\psi^B_{PsH}(1)\right)$. In this wavefunction, $\psi_{PsH}$ denotes a fully correlated atomic wavefunction of $PsH$ which has been optimized by the variational quantum Monte Carlo (VMC) method, with superscripts $A$ and $B$ referring to the position of the two fixed protons, and $S_{AB}$ stands for the usual overlap integrals of the atomic wavefunctions.[1,2] The resulting PEC computed at VMC/HL level (Figure 1 and Table S3 in the SI), is again flat near $R_e$, 8.0 Bohr, with a small BDE, ~3 kJ.mol$^{-1}$. When compared with errors of the HL wavefunction for the hydrogen molecule,[1,2] i.e. almost 17% longer $R_e$ and recovering 66% of the exact BDE, the errors of VMC/HL for $(PsH)_2$ are much larger: 33% longer $R_e$ and recovering only 11% of the exact BDE. No meaningful improvement is observed compared to the MC-HF results, even though "intraatomic" correlations are explicitly included in the VMC/HL wavefunction through the two-particle Jastrow factors,[19] and in addition, the positrons are allowed to be exchanged between the two atoms. In the valence bond theory, the "exchange/resonance" phenomenon is the fundamental origin of covalent bonding in homonuclear diatomic molecules composed of pure matter (and, by symmetry, pure antimatter).[37,38] However, this mechanism proves inefficient to account for bonding in $(PsH)_2$. This conclusion is consistent with our previous findings, which ruled out any significant electron- or positron-associated exchange energy contributions to the BDE based on the TC-IQA analysis of the MC-HF wavefunctions.[22] In the case of the hydrogen molecule, refinements of the HL wavefunction are achieved through the inclusion of extra variational parameters and by accounting for the deformations of atomic wavefunctions in the presence of each other.[39,40] For $(PsH)_2$, this flexibility corresponds to the inclusion of additional two-particle Jastrow factors and their associated variational parameters both in the VMC and DMC wavefunctions, explicitly accounting for the "interatomic" correlations. The corresponding correlation energy, $\Delta E^{corr}(R) = E^{corr}_{(PsH)_2}(R) - 2 \times E^{corr}_{PsH}$, serves as an energetic measure of these correlations (Figure 1 and Table S4 in the SI). Around $R_e$, $\Delta E^{corr}$ accounts for only about 4-6% of the total correlation energy of $(PsH)_2$, and this fraction further decreases at larger inter-proton distances. Nevertheless, near $R_e$, the DEs are primarily governed by the interatomic correlations, whereas at larger separations, $\Delta E^{corr}(R)$ becomes essentially the sole contribution to DEs. Evidently, this relatively small energy component is the exclusive origin of the two-positron gluic bond in $(PsH)_2$ and also dominates the interaction between $PsH$ atoms at inter-proton distances far from equilibrium. For a positronic system, the total correlation energy and the associated $\Delta E^{corr}(R)$ originate from electron-electron ($E^{corr}_{ee}(R)$), electron-positron ($E^{corr}_{ep}(R)$), and positron-positron ($E^{corr}_{pp}(R)$) correlations. Ideally, one would like to disentangle these contributions; however, this task is not straightforward, since different correlation partitioning schemes exist,[41–44] and in some of them, the total correlation energy is not strictly additive, $E^{corr}(R) \neq E^{corr}_{ee}(R) + E^{corr}_{ep}(R) + E^{corr}_{pp}(R)$

.[45,46] The simplest correlated computational method that preserves additivity for MC systems is second-order many-body perturbation theory.[47] This approach, an extension of the conventional second-order Møller–Plesset perturbation theory to MC quantum systems (MC-MP2), employs the MC-HF Hamiltonian as the zeroth-order reference.[48–50]

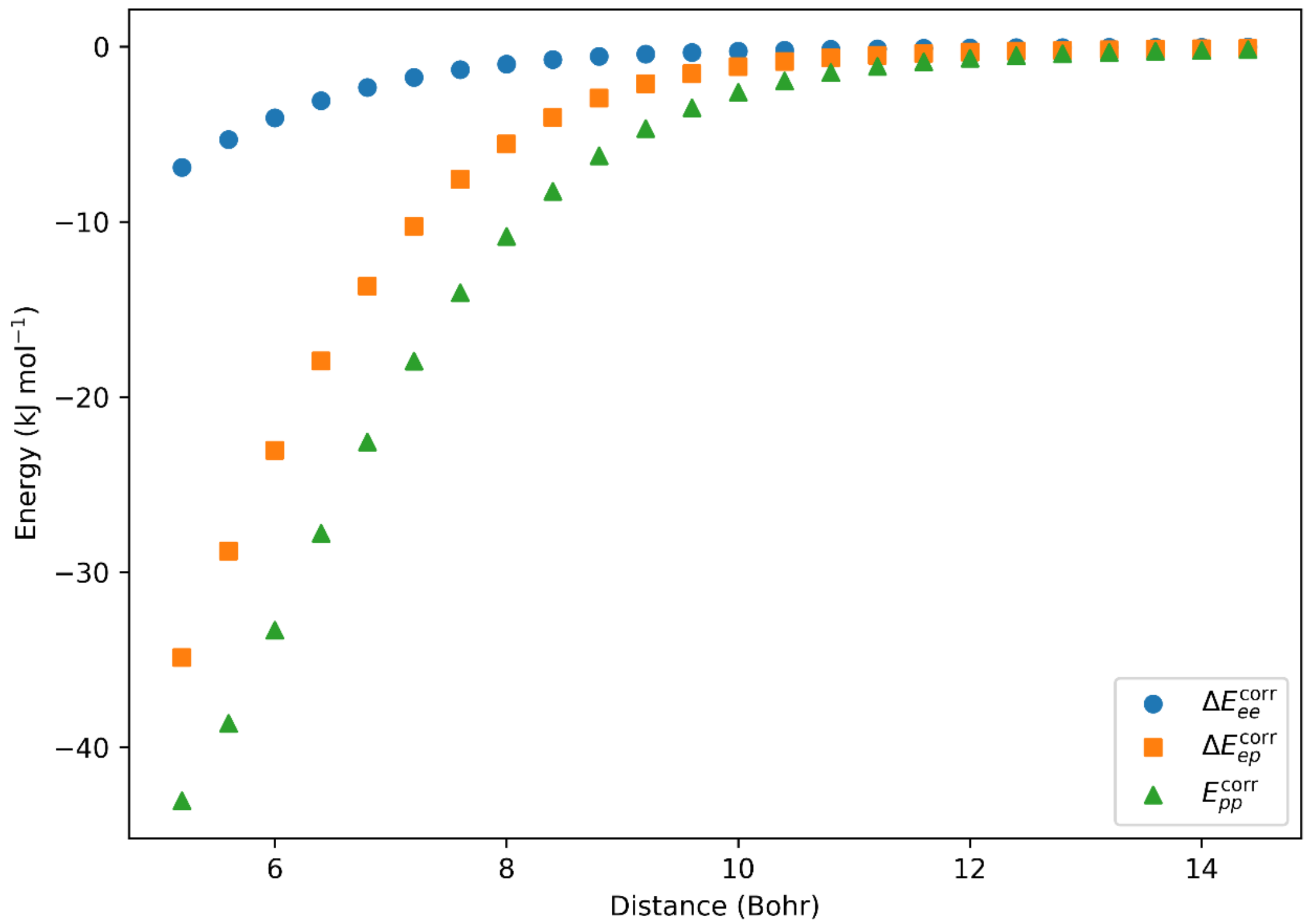

**Figure 2**- The three contributions — electron-electron, electron-positron, and positron-positron — to the interatomic correlation energy of $(PsH)_2$, evaluated at the MC-MP2 level as functions of the inter-proton distance (see the text for definitions and abbreviations).

Although the correlation energies obtained at this level are less accurate than those derived from quantum Monte Carlo methods, they are used herein to provide qualitative insight into the relative significance of the different types of correlations. Among the three contributions to the interatomic correlation energy evaluated at the MC-MP2 level, $\Delta E_{MC-MP2}^{corr}(R) = \Delta E_{ee}^{corr}(R) + \Delta E_{ep}^{corr}(R) + E_{pp}^{corr}(R)$ (Figure 2 and Table S5 in the SI), $\Delta E_{ee}^{corr}$ is the least significant, contributing less than 10% of $\Delta E_{MC-MP2}^{corr}$ at the entire range of the considered inter-proton distances (5.2 - 14.4 Bohr). In contrast, the two dominant contributions arise from $\Delta E_{ep}^{corr}(R)$ and $E_{pp}^{corr}(R)$, which account for approximately 30-40% and 50-65% of $\Delta E_{MC-MP2}^{corr}(R)$, respectively. Manifestly, these two "quantum" components constitute the principal driving force behind the formation of the two-positron gluic bond in $(PsH)_2$. Now, the key question then arises as to whether this bond can be classified within the known categories of chemical bonds or interatomic interactions.

For systems composed of pure matter (and, by symmetry, from pure antimatter), the only interactions that emerge purely from quantum correlations among electrons (and, by symmetry, among positrons) are dispersion (London) interactions.[51–53] The perturbative treatment of the dispersion between two charge-neutral atoms, $A$ and $B$, leads to the asymptotic expansion: $E_{disp}^{AB} = -\frac{C_6^{AB}}{R^6} - \frac{C_8^{AB}}{R^8} - \frac{C_{10}^{AB}}{R^{10}} - ...$, which is convergent at the long inter-nuclear, $R$, separations where $C_n^{AB}$ are the dispersion coefficients.[53] For the interaction between two $PsH$ atoms, to the best of the authors' knowledge, the only reported dispersion coefficients are those of Yan, $C_6^{PsH-PsH} = 273.66$ and $C_8^{PsH-PsH} = 24554$ (in atomic units).[54] The corresponding $E_{disp}^{PsH-PsH}$, constructed from these coefficients, is clearly much less stabilizing than the actual DMC computed PEC for $R > 6.8$ Bohr (see the blue curve in Figure 1). Even at the largest inter-proton distances considered, no indication of an asymptotic convergence of the PEC to $E_{disp}$ is observed. To elucidate the origin of this discrepancy, at first step the $C_6^{PsH-PsH}$ value was rechecked through an independent estimation using Tang's interpolation formula: $C_6^{AB} = \frac{2\alpha_A \alpha_B C_6^{AA} C_6^{BB}}{(\alpha_A)^2 C_6^{BB} + (\alpha_B)^2 C_6^{AA}}$,[55] and employing the much better established $C_6$ values for the interaction of two hydrogen atoms,[56] and a hydrogen and a $PsH$ atom,[54,57] ( $A=H$, $B=PsH$), together with the static polarizabilities of the atoms, $\alpha_{A/B}$. Using the literature values, $C_6^{H-H}=6.5$, $C_6^{H-PsH}=40.3$, $\alpha_H=4.5$, $\alpha_{PsH}=42.3$ (all in atomic units), a value of $C_6^{PsH-PsH}=282.5$ is obtained, which is only 3% larger than Yan's result. Alternatively, the Slater-Kirkwood equation: $C_6^{AB}=\left(\frac{3}{2}\right)\frac{\alpha_A\alpha_B}{\sqrt{\frac{\alpha_A}{N_A}}+\sqrt{\frac{\alpha_B}{N_B}}}$,[58,59] where $N_{A/B}=N_{A/B}^{e}-N_{A/B}^{p}$ ($N_{A/B}^{e}/N_{A/B}^{p}$ is the number of electrons/positrons of atom $A/B$), was used to estimate the dispersion coefficient, yielding $C_6^{PsH-PsH}=\left(\frac{3}{4}\right)\alpha_{PsH}\sqrt{\alpha_{PsH}N_{PsH}}=291.6$ ($\alpha_{PsH}=42.3, N_{PsH}=2$), again differing from Yan's value by less than 10%. These close agreements rule out any significant numerical error in the reported value of $C_6^{PsH-PsH}$. A more plausible explanation for the observed discrepancy is therefore the non-negligible contribution of the higher-order terms in the dispersion series, such as $-\frac{C_{10}}{R^{10}}$, even at inter-proton distances substantially larger than $R_e$. Since to the best of the authors' knowledge, no values for $C_n, n\geq 10$ have yet been reported for $PsH-PsH$ interactions, this hypothesis must be examined in future studies. Nevertheless, the unusually large ratio of dispersion coefficients, $\frac{C_8^{PsH-PsH}}{C_6^{PsH-PsH}}\sim 90$, compared with the corresponding ratio for ordinary few-electron atoms,[53] already points to the enhanced importance of higher-order dispersion coefficients in the asymptotic expansion. In our previous study,[22] we compared the bond in $(PsH)_2$ with that in $Li_2$. However, among few-electron ordinary atoms and constituent charged-neutral dimers, the static polarizability of $Be$ atom, $\alpha_{Be}=37.7$,[60] the dispersion coefficients of $Be_2$, $C_6^{Be-Be}=214$ and $C_8^{Be-Be}=10230$ (all in atomic units),[60,61] and its BDE, ~11.2 kJ.mol$^{-1}$,[61] are in fact closest to those of $PsH$ and $(PsH)_2$. Interestingly, $Be-Be$ interaction in $Be_2$ is one of the most extensively studied,[62–68] and, at the same time, among the most intricate and controversial bonds formed between any two few-electron atoms; similar to $(PsH)_2$, the PEC of $Be_2$ has no genuine minimum at HF level.[69–71] Since $Be_2$ consists of two closed-shell $Be$ atoms, the usual electron-exchange mechanics is also ineffective in the bond formation of this system. Instead, its bonding arises from a subtle interplay between strong dispersion interactions and configuration mixing between ground and certain excited electronic states.[62–68] Given that the dispersion coefficients of $(PsH)_2$ system, as well as its BDE, are even larger than $Be_2$, we conclude that $(PsH)_2$ and its bond may appropriately be characterized as a "super" van der Waals complex and a "super" van der Waals bond.

## Conclusion

The present study demonstrates that the nature of the two-positron gluic bond is fundamentally distinct from that of its one-positron counterpart, thereby revealing a novel bonding mechanism. This finding, however, likely represents only the beginning of a broader class of phenomena. In a recent work,[24] Reyes and coworkers proposed a unique "positron-driven" bonding mechanism operating upon the attachment of a positron to $Be_2$. Moreover, the existence of a three-positron bond with a BDE comparable to that of $(PsH)_2$ has been established through the formation of $e^+(PsH)_2$,[20] obtained by adding a positron to $(PsH)_2$, although the precise nature of this three-positron bond remains to be clarified. The possibility of three-center two-positron bonds has also been put forward,[21] together with speculative connections to antimatter aromaticity, yet a detailed investigation is still required to elucidate the underlying bonding mechanism. Taken together, these observations suggest that bonding phenomena at the interface of matter and antimatter constitute a largely unexplored and potentially rich domain, promising fundamentally new concepts and unexpected forms of interactions.

**Computational details**

The many-particle Schrödinger equation was solved, treating four electrons and two positrons as quantum particles that interact with each other and with two fixed point-charge protons via Coulomb's law. The details of the variational methods used to obtain the ground-state total energy at each inter-proton distance, i.e., the potential energy curve of $(PsH)_2$, have been fully described in previous studies,[19,22]; therefore, only a brief account is provided here.

The algebraic multi-component Hartree-Fock equations were solved using a self-consistent field procedure with a convergence threshold of at least $10^{-8}$ hartree.[32] The nucleus-centered aug-cc-pVQZ Gaussian basis set, originally developed for the hydrogen atom,[33,34] was employed to expand both electronic and positronic spatial orbitals and the method is denoted as MC-HF/[QZ:QZ]. The resulting spin-singlet closed-shell ground state is represented by an energy-optimized wavefunction, constructed as the product of 4×4 and 2×2 Slater determinants for electrons and positrons, respectively. At the next step, employing the formalism of the MC-Møller–Plesset many-body perturbation theory and the spin-orbitals derived from MC-HF/[QZ:QZ] level,[50] the following equations were used to compute the electron-electron, electron-positron, and positron-positron contributions to the total energy correlation energy: $E_{ee}^{corr}=\left(\frac{1}{4}\right)\sum_{abrs}\frac{\left|\langle ab\|rs\rangle\right|^2}{\varepsilon_a^e+\varepsilon_b^e-\varepsilon_r^e-\varepsilon_s^e}$,

$E_{ep}^{corr}=\sum_{aa'rr'}\frac{\left|\langle aa'|rr'\rangle\right|^2}{\varepsilon_a^e+\varepsilon_{a'}^p-\varepsilon_r^e-\varepsilon_{r'}^p}$, $E_{pp}^{corr}=\left(\frac{1}{4}\right)\sum_{a'b'r's'}\frac{\left|\langle a'b'\|r's'\rangle\right|^2}{\varepsilon_{a'}^p+\varepsilon_{b'}^p-\varepsilon_{r'}^p-\varepsilon_{s'}^p}$, where unprimed/primed indices ($a$, $b$) denote occupied, and ($r$, $s$) virtual, electronic/positronic spin-orbitals; $\langle ij\|kl\rangle=\langle ij\,|\,kl\rangle-\langle ij\,|\,lk\rangle$, where $\langle ij\,|\,kl\rangle/\langle i'j'\,|\,k'l'\rangle$ are standard two-electron/positron

integrals while $\langle ij' | kl' \rangle$ are the newly introduced one-electron-one-positron integrals;[50] and $\varepsilon$ refers to electronic/positronic orbital energies. To obtain the best energy-optimized potential energy curve, fully correlated variational and diffusion quantum Monte Carlo methods in the fixed-node approximation were applied.[29–31,72] The trial spin-singlet wavefunction included anti-symmetrized two-particle Jastrow factors for each pair of particles with multiple variational parameters, whose explicit form has been given previously.[19,22] In a multi-step procedure, the variational parameters of the wavefunction were first optimized via variational quantum Monte Carlo by minimizing the mean absolute deviation of the local energy,[29] and subsequently refined through the total energy minimization for the inter-proton distances considered. Finally, the fully optimized variational quantum Monte Carlo wavefunction was employed as input in the fixed-node diffusion quantum Monte Carlo simulations, performed with 3000 walkers and a time step of 0.003 hartree$^{-1}$. The diffusion quantum Monte Carlo-derived potential energy curve provides an upper bound to the exact potential energy curve and is considered the most accurate result obtained in this study. The primary limitation in potential energy curve accuracy arises from the statistical noise inherent in the quantum Monte Carlo sampling procedures, leading to estimated energy errors in the range of $10^{-4} - 10^{-5}$ Hartree across all the considered inter-proton distances. To ensure the reliability, total energies are therefore reported only to four decimal places in Hartree. The zero-point energies have been computed by solving the one-dimensional nuclear Schrödinger equation for the diffusion quantum Monte Carlo-derived potential energy curves using the Numerov method, as detailed previously.[22]

## Authors contributions

Dr. Mohammad Goli: Conceptualization, Data curation, Formal analysis, Investigation, Methodology, Software, Validation, Writing – review & editing.

Dr. Dario Bressanini: Conceptualization, Data curation, Formal analysis, Investigation, Methodology, Software, Validation, Writing – review & editing.

Dr. Shant Shahbazian: Conceptualization, Funding acquisition, Formal analysis, Investigation, Methodology, Validation, Writing – original draft, Writing – review & editing.

## Conflicts of interest

There are no conflicts to declare.

## Data availability

The data supporting this article have been included as part of the Supplementary Information. Supplementary information: Tables S1 to S5.

## Acknowledgements

This work has been supported by HESCS of Armenia, grant No. 10-27/25IRF. The authors are grateful to Prof. W. H. Eugen Schwarz for his helpful comments.

## Notes and References

# Supporting Information

# The two-positron gluic bond as a manifestation of 'super' van der Waals interactions

Mohammad Goli,[a] Dario Bressanini,[b] and Shant Shahbazian[c]

[a]Institute of Physics, Faculty of Physics, Astronomy, and Informatics, Nicolaus Copernicus University in Toruń, ul. Grudziądzka 5, 87-100 Toruń, Poland,
E-mail: mgoli@umk.pl

[b]Dipartimento di Scienza e Alta Tecnologia, Università dell'Insubria, Como, Italy
E-mail: dario.bressanini@uninsubria.it

[c][Alikhanian National Laboratory (Yerevan Physics Institute), Alikhanian Brothers Street 2, Yerevan 0036, Armenia
E-mail: shant.shahbazyan@aanl.am

**Table S1** Total energies of $\left[H^-, e^+, H^-\right]$ as a function of the inter-proton distance computed at the DMC and MC-HF/[QZ:QZ] levels of theory for the singlet electronic spin states. The statistical errors of the QMC-derived energies are always below $10^{-4}$ Hartree. All values are given in atomic units.[*]

| $R_{HH}$ | DMC | MC-HF | $R_{HH}$ | DMC | MC-HF |
|---|---|---|---|---|---|
| 3.2 | -1.3107 | -1.1127 | 8.4 | -1.3351 | -1.1736 |
| 3.6 | -1.3165 | -1.1248 | 8.6 | -1.3344 | -1.1726 |
| 4.0 | -1.3229 | -1.1370 | 8.8 | -1.3333 | -1.1715 |
| 4.4 | -1.3290 | -1.1478 | 9.0 | -1.3324 | -1.1704 |
| 4.8 | -1.3337 | -1.1567 | 9.2 | -1.3317 | -1.1693 |
| 5.2 | -1.3370 | -1.1637 | 9.4 | -1.3310 | -1.1682 |
| 5.6 | -1.3392 | -1.1690 | 9.6 | -1.3301 | -1.1670 |
| 5.9 | -1.3399 | -1.1720 | 9.8 | -1.3294 | -1.1659 |
| 6.0 | -1.3401 | -1.1728 | 10.0 | -1.3286 | -1.1648 |
| 6.1 | -1.3403 | -1.1735 | 10.2 | -1.3280 | -1.1638 |
| 6.2 | -1.3404 | -1.1742 | 10.4 | -1.3271 | -1.1627 |
| 6.3 | *-1.3405* | -1.1747 | 10.6 | -1.3266 | -1.1618 |
| 6.4 | -1.3404 | -1.1752 | 10.8 | -1.3258 | -1.1609 |
| 6.5 | -1.3404 | -1.1756 | 11.0 | -1.3253 | -1.1601 |
| 6.6 | -1.3404 | -1.1760 | 11.2 | -1.3246 | -1.1594 |
| 6.7 | -1.3402 | -1.1763 | 11.4 | -1.3242 | -1.1587 |
| 6.8 | -1.3401 | -1.1765 | 11.6 | -1.3235 | -1.1582 |
| 6.9 | -1.3398 | -1.1767 | 11.8 | -1.3231 | -1.1578 |
| 7.2 | -1.3392 | *-1.1769* | 12.0 | -1.3224 | -1.1574 |
| 7.6 | -1.3379 | -1.1764 | 12.2 | -1.3220 | -1.1570 |
| 8.0 | -1.3365 | -1.1753 | 12.4 | -1.3217 | -1.1567 |
| 8.2 | -1.3357 | -1.1745 | 12.6 | -1.3213 | -1.1564 |

[*]To compute the bond dissociation energies from the total energies, the following data have been employed: For $PsH$, DMC and MC-HF/[QZ:QZ] total energies are -0.78920 and -0.66617 Hartree, respectively, whereas for $H^-$, DMC and MC-HF/[QZ:QZ] total energies are -0.52775 and -0.48781 Hartree, respectively.

**Table S2** Total energies of $(PsH)_2$ as a function of the inter-proton distance computed at the DMC and MC-HF/[QZ:QZ] levels of theory for the singlet electronic and positronic spin states. The statistical errors of the QMC-derived energies are always below $10^{-4}$ Hartree. All values are given in atomic units.[*]

| $R_{HH}$ | DMC | MC-HF | $R_{HH}$ | DMC | MC-HF |
|---|---|---|---|---|---|
| 3.6 | -1.5584 | -1.2890 | 9.6 | -1.5808 | -1.3326 |
| 4.0 | -1.5695 | -1.3009 | 9.8 | -1.5804 | -- |
| 4.4 | -1.5782 | -1.3104 | 10.0 | -1.5802 | -1.3326 |
| 4.8 | -1.5837 | -1.3177 | 10.2 | -1.5799 | -- |
| 5.2 | -1.5867 | -1.3230 | 10.4 | -1.5798 | -1.3325 |
| 5.6 | -1.5884 | -1.3267 | 10.6 | -1.5796 | -- |
| 5.8 | -1.5888 | -- | 10.8 | -1.5794 | -1.3325 |
| 6.0 | *-1.5888* | -1.3292 | 11.0 | -1.5793 | -- |
| 6.2 | -1.5885 | -- | 11.2 | -1.5792 | -1.3325 |
| 6.4 | -1.5883 | -1.3308 | 11.4 | -1.5791 | -- |
| 6.6 | -1.5880 | -- | 11.6 | -1.5790 | -1.3325 |
| 6.8 | -1.5875 | -1.3318 | 11.8 | -1.5789 | -- |
| 7.0 | -1.5870 | -- | 12.0 | -1.5788 | -1.3325 |
| 7.2 | -1.5865 | -1.3323 | 12.2 | -1.5788 | -- |
| 7.4 | -1.5859 | -- | 12.4 | -1.5788 | -- |
| 7.6 | -1.5854 | -1.3326 | 12.6 | -1.5787 | -- |
| 7.8 | -1.5847 | -- | 12.8 | -1.5786 | -- |
| 8.0 | -1.5841 | -1.3327 | 13.0 | -1.5786 | -- |
| 8.2 | -1.5836 | -1.3327 | 13.2 | -1.5786 | -- |
| 8.3 | -- | *-1.3327* | 13.4 | -1.5786 | -- |
| 8.4 | -1.5831 | -1.3327 | 13.6 | -1.5785 | -- |
| 8.6 | -1.5826 | -1.3327 | 13.8 | -1.5785 | -- |
| 8.8 | -1.5822 | -1.3327 | 14.0 | -1.5785 | -- |
| 9.0 | -1.5817 | -- | 14.2 | -1.5785 | -- |
| 9.2 | -1.5814 | -1.3327 | 14.4 | -1.5785 | -- |
| 9.4 | -1.5810 | -- | | | |

[*] To compute the bond dissociation energies from the total energies, the following data have been employed for $PsH$: DMC and MC-HF/[QZ:QZ] total energies are -0.78920 and -0.66617 Hartree, respectively.

**Table S3** Total energies of $(PsH)_2$ computed using the VMC method, and employing Heitler and London's wavefunction, denoted as VMC/HL (see the main text for details), as a function of the inter-proton distance, for the singlet electronic and positronic spin states. The statistical errors of the VMC-derived energies are always equal to or below $10^{-4}$ Hartree. All values are given in atomic units.

| $R_{HH}$ | VMC | $R_{HH}$ | VMC |
|---|---|---|---|
| 4.0 | -1.5288 | 9.6 | -1.5732 |
| 4.4 | -1.5448 | 10.0 | -1.5731 |
| 4.8 | -1.5559 | 10.4 | -1.5730 |
| 5.2 | -1.5630 | 10.8 | -1.5730 |
| 5.6 | -1.5676 | 11.2 | -1.5730 |
| 6.0 | -1.5704 | 11.6 | -1.5729 |
| 6.4 | -1.5722 | 12.0 | -1.5729 |
| 6.8 | -1.5732 | 12.4 | -1.5729 |
| 7.2 | -1.5737 | 12.8 | -1.5728 |
| 7.6 | -1.5737 | 13.2 | -1.5728 |
| 8.0 | *-1.5738* | 13.6 | -1.5728 |
| 8.4 | -1.5735 | 14.0 | -1.5727 |
| 8.8 | -1.5735 | 14.4 | -1.5727 |
| 9.2 | -1.5734 | 14.8 | -1.5727 |

$^{*}$ To compute the bond dissociation energies from the total energies, the following data have been employed $PsH$: the VMC total energy is -0.78630 Hartree.

**Table S4** The interatomic correlation energies (see the main text for definition) computed for the ground state $(PsH)_2$ as a function of the inter-proton distance. All values are given in atomic units.*

| $R_{HH}$ | $\Delta E^{corr}(R)$ | $R_{HH}$ | $\Delta E^{corr}(R)$ |
|---|---|---|---|
| 6.0 | -0.0135 | 9.2 | -0.0027 |
| 6.4 | -0.0114 | 9.6 | -0.0021 |
| 6.8 | -0.0097 | 10.0 | -0.0016 |
| 7.2 | -0.0081 | 10.4 | -0.0012 |
| 7.6 | -0.0067 | 10.8 | -0.0009 |
| 8.0 | -0.0054 | 11.2 | -0.0006 |
| 8.4 | -0.0043 | 11.6 | -0.0005 |
| 8.8 | -0.0034 | 12.0 | -0.0003 |

* The employed correlation energy of $PsH$ is 0.12302 Hartree.

**Table S5** The three contributions to the interatomic correlation energy computed at the MC-MP2 level (see the main text for details), for the ground state of $(PsH)_2$ as a function of the inter-proton distance. All values are given in atomic units.*

| $R_{HH}$ | $\Delta E_{ee}^{corr}(R)$ | $\Delta E_{ep}^{corr}(R)$ | $E_{pp}^{corr}(R)$ |
|---|---|---|---|
| 5.2 | -0.0026 | -0.0133 | -0.0164 |
| 5.6 | -0.0020 | -0.0110 | -0.0147 |
| 6.0 | -0.0015 | -0.0088 | -0.0127 |
| 6.4 | -0.0012 | -0.0068 | -0.0106 |
| 6.8 | -0.0009 | -0.0052 | -0.0086 |
| 7.2 | -0.0007 | -0.0039 | -0.0068 |
| 7.6 | -0.0005 | -0.0029 | -0.0053 |
| 8.0 | -0.0004 | -0.0021 | -0.0041 |
| 8.4 | -0.0003 | -0.0015 | -0.0031 |
| 8.8 | -0.0002 | -0.0011 | -0.0024 |
| 9.2 | -0.0002 | -0.0008 | -0.0018 |
| 9.6 | -0.0001 | -0.0006 | -0.0013 |
| 10.0 | -0.0001 | -0.0004 | -0.0010 |
| 10.4 | -0.0001 | -0.0003 | -0.0007 |
| 10.8 | -0.0001 | -0.0002 | -0.0006 |
| 11.2 | 0.0000 | -0.0002 | -0.0004 |
| 11.6 | 0.0000 | -0.0002 | -0.0003 |
| 12.0 | 0.0000 | -0.0001 | -0.0003 |
| 12.4 | 0.0000 | -0.0001 | -0.0002 |
| 12.8 | 0.0000 | -0.0001 | -0.0002 |
| 13.2 | 0.0000 | -0.0001 | -0.0001 |
| 13.6 | 0.0000 | -0.0001 | -0.0001 |
| 14.0 | 0.0000 | -0.0001 | -0.0001 |
| 14.4 | 0.0000 | 0.0000 | -0.0001 |

* The used electron-electron and electron-positron correlation energies of $PsH$ are 0.02884 and 0.03633 Hartree, respectively.